\documentclass[preprint]{elsarticle}

\newcommand{\be}{\begin{equation}}
\newcommand{\en}{\end{equation}}
\newcommand{\bea}{\begin{eqnarray}}
\newcommand{\ena}{\end{eqnarray}}
\usepackage{xcolor}
\usepackage[normalem]{ulem}
\usepackage{graphicx}
\usepackage{soul} 

\journal{Physica A}

\begin{document}
\begin{frontmatter}

\title{Complementarity relation for irreversible processes near steady states}

\author[cnen,ungs,cbpf]{E.S. Santini\corref{cor1}}
\ead{santini@cbpf.br}

\author[ungs,conicet]{M.F. Carusela}
\ead{flor@ungs.edu.ar}

\author[ubacbc,ubavet]{E.D. Izquierdo}
\ead{edu@ungs.edu.ar}

\cortext[cor1]{Corresponding author}


\address[cnen]{Comiss\~ao Nacional de Energia Nuclear \\ 
Rua General Severiano 90, Botafogo (22290-901) -- Rio de Janeiro, RJ -- Brazil}
\address[ungs]{Universidad Nacional de General Sarmiento, Instituto de Ciencias \\ 
  J. M. Guti\'errez 1150 -- Malvinas Argentinas (1613)-- Pcia de Buenos Aires -- Argentina }
\address[cbpf]{
Centro Brasileiro de Pesquisas F\'{\i}sicas - ICRA-BR \\
Rua Dr. Xavier Sigaud 150, Urca (22290-180) -- Rio de Janeiro, RJ -- Brazil}
\address[ubacbc]{Ciclo B\'asico Com\'un, Universidad de Buenos Aires \\ Cdad. Universitaria, Pab. 3, (1428) Buenos Aires -- Argentina.}
\address[ubavet]{ Facultad de Agronom\'ia \\ Universidad de Buenos Aires, 
Av. San Martín 4453, Cap. Fed., Buenos Aires (C1417DSE) -- Argentina.}
\address[conicet]{Consejo Nacional de Investigaciones Cient\'{\i}ficas y T\'ecnicas \\  Av. Rivadavia 1917, (1033) -- Buenos Aires -- Argentina}

\begin{abstract}
A relation giving a minimum for the  irreversible work in quasi-equilibrium processes was derived by Sekimoto et al. (K. Sekimoto and S. Sasa, J. Phys. Soc. Jpn. {\bf 66} (1997), 3326) in the framework of stochastic energetics. This relation can also be written as a type of ``uncertainty principle'' in such a way that the precise determination of the Helmholtz free energy through the observation of the work $<W>$ requires an indefinitely large experimental time $\Delta t$. In the present article, we extend this relation to the case of quasi-steady processes by using the concept of non-equilibrium Helmholtz free energy. We give a formulation of the second law for these processes that extends that presented by Sekimoto (K. Sekimoto, Prog. Theo. Phys. Suppl. No. {\bf 130} (1998), 17) by  a term of the first order in the inverse of the experimental time. As application of our results, two possible experimental situations are considered: stretching of a RNA molecule and the drag of a dipolar particle in the presence of a gradient of electric force.
\end{abstract}



\end{frontmatter}

\section{Introduction}
In recent years, there has been a growing interest in studying small mesoscopic systems, immersed in different substrates, such as colloidal particles, nanoparticles in solutions, or biological systems, all of which are dominated by fluctuations. The principal interest is motivated due to recent experimental breakthroughs and technical applications.
Thermodynamic notions, such as applied work, dissipated heat and entropy, have been used successfully to analyze processes, in which single colloidal particles or biomolecules are manipulated externally \cite{bustamante}. In this area, it is possible to find several studies that have focused on the generation of non-equilibrium situations, from a time-dependent potential, manipulated externally to model the effect of moving laser traps, micropippetes, or atomic force microscopic tips. In all these cases, it is straightforward to find a clear identification of external work, internal energy and dissipated heat, whose consistency has been proven in many experiments, going from the micro to the nano world  \cite{wang, liphardt, collin, blickle, douarche, Jarzinsky2008, Toyabel}.

Several authors have also established a connection between the phenomena related to non-equilibrium steady states and the thermodynamic laws for slow processes connecting different steady states\cite{ken1, jou, keizer, eu, hatano, oono}. 
The main objective of this paper is to shed some light on this last point, by addressing the analysis of a quasi-steady process, consisting of a particle in contact with a single
bath at temperature T, while some control parameter changes slowly. 

The dynamics of this particle can be modeled by a Langevin equation in the framework of Stochastic Energetics (SE)(see below).

According to thermodynamics, if we consider a system in contact with a heat bath and control parameters changes quasi-statically, the work $W$ (done on system) needed for the change is equal to the variation of Helmholtz free energy, $\Delta F$:
\be
W=\Delta F \; , 
\en

being $\Delta F$ composed by the sum of the reversible heat released to the heat bath, $Q_{rev}$ and the change of internal energy, $\Delta E$: $ \Delta F=Q_{rev}+\Delta E$. If the change of the control parameters is not quasi-static, then the necessary work $W$ is larger than the reversible one

\be 
 W- \Delta F = Q_{irr}\geq 0 \; , 
\en

where $Q_{irr}$ is the irreversible heat that is equal to the difference between the released heat $Q$ and the reversible heat: $Q_{irr}\equiv Q- Q_{rev}$. The released heat $ Q$ satisfies the first law : 
 
 \be
 Q+\Delta E=W \;. 
 \en

In this definition we have adopted that  if $Q>0$  the system dissipates energy to the enviroment.

 In order to obtain the released heat $Q$ for a given protocol of control parameters, it is necessary to have both a dynamical model of the system and a kinematical interpretation of the heat released by the system. An approach was introduced to obtain $Q$ for systems whose dynamics are decribed by Langevin equations which are now known as {\it stochastic energetics} \cite{ken, kenbook, Seif}. It constitutes an intermediary level of description that lies between Hamiltonian dynamics including all degrees of freedom of the concerned system, and thermodynamics where the system is controlled by external agents. In the framework of this approach, for a system that follows a quasi-static  isothermal  process, a complementarity relation giving a minimum for the product of the irreversible heat times the experimental time, $Q_{irr}\Delta t \geq k_{B}T{\cal S}_{min}$, was demontrated in \cite{ken1} and an expression for the second law with a first order correction was obtained in \cite{ken2}. An extension of the 
stochastic energetics to  the case of quasi-steady (QS) processes and an expression for the second law to zero {\it{-th}} order were presented in \cite{ken2}. In the present letter we continue, in one sense, the work initiated in \cite{ken2} by generalizing the approach used in \cite{ken1}. We are able to show a complementarity relation that is valid for QS processes together with an expression for the second principle with a first order correction. We propose two possible experimental situations in order to discuss the relevance of our results. They are the stretching of a RNA molecule and the drag of a dipolar particle subjected  to a gradient of electric field. For the last we detect a difference in the irreversible heat given by a temperature-dependent term, which appears only in one case, i.e. when the stiffness constant is varied too.

This work is organized as follows: in the next section we apply the approach of SE to the case of the QS processes followed by a system satisfying a Langevin equation (already used in \cite{ken2} but with slightly more general expression for the applied forces). At the same time, we sketch the principal steps of SE. We obtain our results and then in section \ref{3} we exemplify them by studying a simple model. In sections \ref{4} and \ref{5} we illustrate the results in two models of possible applications related to the stretching of a RNA molecule and to the drag of a dipolar particle in the presence of a electric gradient. The next section is for the conclusions. Some computing is included in the appendix, in order not to deviate the text from the principal line of reasoning.

\section{Stochastic energetics. The case of irreversible processes near steady states. The second principle}
\label{2}

We are going to extend the approach followed in \cite{ken1} for irreversible processes near equilibrium states, to the case of irreversible processes near steady states.
To that end we consider a generalization of the model studied in \cite{ken2} section 5, which exhibes the features of a QS process. We study the response of a system in contact with a single bath, at temperature $T$, in the limit of very strong friction.
Let $\mathbf{x}=\{x_1,...x_n\}$ representing the state of the fluctuating system and $\mathbf{b}=\{b_1,...b_r\}$  the parameters that control the system through the potential $U(\mathbf{x}(t);\mathbf{a}(t); \mathbf{b}(t))=U(\mathbf{x}(t)-\mathbf{a}(t); \mathbf{b}(t))$. The quantity $\mathbf{a}(t)$ is another parameter that models a conservative force. We consider the QS state where $\dot{a}=\mathbf{v}$, with $\mathbf{v}$ a constant velocity and, for simplicity, we can take $a(0)=0$.
Besides the conservative forces arising from the potential $U(\mathbf{x}-\mathbf{a}; \mathbf{b}(t))$, we can perturb the particle by a direct force $\mathbf{f}$ which we assume constant. This force may include all the conservative and non-conservative contributions that do not depend on $\mathbf{a}$.
We describe the stochastic particle motion through the Langevin equation as follows:

\be \label{LsG}
\Gamma \cdot \frac{d\mathbf{x}}{dt}=-\frac{\partial U}{\partial \mathbf{x}}(\mathbf{x};\mathbf{a}; \mathbf{b})+\mathbf{\xi}(t)+\mathbf{f}
\en

where $\Gamma$ is a friction constant given by a symmetric and positive definite matrix,  and  $\mathbf{\xi}(t)$ is a Gaussian and white-correlated stochastic force, satisfying

\be
\left\langle \mathbf{\xi}(t) \right\rangle=0\ ,\  \left\langle \mathbf{\xi}(t)^{t}\mathbf{\xi}(t')\right\rangle=2\frac{\Gamma}{\beta}\delta(t-t').
\en 
where $\beta \equiv \frac{1}{k_{B}T}$.
In close analogy with the case of quasi-equilibrium process studied in \cite{ken1},  we rewrite Eq.(\ref{LsG}) by making the scalar product\footnote{The multiplication of fluctuating quantities, i.e. $\mathbf{\xi}(t)\cdot d\mathbf{x}$, should be understood in the sense of Stratonovich calculus \cite{gardiner}.} by $d\mathbf{x}$ along the {\it realized} trajectory and  using that $dU= \frac{\partial U}{\partial \mathbf{x}}.d\mathbf{x}+\frac{\partial U}{\partial \mathbf{b}}d\mathbf{b}+\frac{\partial U}{\partial \mathbf{a}}d\mathbf{a}$, obtaining a balance equation for energy, which is:

\be \label{LesGb}
\left(\Gamma \cdot \frac{d\mathbf{x}}{dt}-\mathbf{\xi}(t)\right)\cdot d\mathbf{x}+dU(\mathbf{x};\mathbf{a};\mathbf{b})-\mathbf{f}\cdot d\mathbf{x}= \frac{\partial U}{\partial \mathbf{b}}\left(\mathbf {x};\mathbf{a};\mathbf{b} \right)\cdot d\mathbf{b}+\frac{\partial U}{\partial \mathbf{a}}\left(\mathbf {x};\mathbf{a};\mathbf{b} \right)\cdot d\mathbf{a}\, \, .
\en

We define an effective energy,
\be 
E(\mathbf{x},\mathbf{a},\mathbf{b})= U(\mathbf{x};\mathbf{a}; \mathbf{b})-\mathbf{f} \cdot \mathbf{x}
\en 
Through $\mathbf{f}$  field, $E$ includes the energy exchanged with the outside.

We can rewrite the expression (\ref{LesGb}) as

\be \label{Les2}
d'Q +dE =d'W \, \, \, \, ,
\en

where 
\be
d'Q=\left(\Gamma \cdot \frac{d\mathbf{x}}{dt}-\mathbf{\xi}(t)\right)\cdot d\mathbf{x}
\en 

is the heat discharged onto the bath and (using that $\frac{\partial U}{\partial \mathbf{a}}=-\frac{\partial U}{\partial \mathbf{x}}$)

\be \label{trabalho}
d'W=\frac{\partial U}{\partial \mathbf{b}}\left(\mathbf{x};\mathbf{a};\mathbf{b} \right)\cdot d\mathbf{b}-\frac{\partial U}{\partial \mathbf{x}}\left(\mathbf{x};\mathbf{a};\mathbf{b} \right)\cdot d\mathbf{a}
\en 
is the total work done by the external agent to the system.

Taking the steady-state average of the Langevin equation (\ref{LsG}) we have

\be \label{LM}
\Gamma \cdot \left\langle \mathbf{v}\right\rangle=\left\langle-\frac{\partial U}{\partial \mathbf{x}}(\mathbf{x};\mathbf{a}; \mathbf{b})\right\rangle +\mathbf{f}
\en 

where we used  that $\left\langle\dot{\mathbf{x}}\right\rangle\equiv\mathbf{v}$ with $\mathbf{v}=constant$

Now we can take the steady-state average of the total work, equation (\ref{trabalho}).Using (\ref{LM}) and using that $\mathbf{a}=\mathbf{v} t$ we obtain:

\be \label{trabalho2}
\left\langle d'W\right\rangle=\left\langle \frac{\partial U}{\partial \mathbf{b}}\left(\mathbf{x};\mathbf{a};\mathbf{b} \right)\right\rangle\cdot d\mathbf{b}+\left\langle -\frac{\partial U}{\partial \mathbf{x}}\left(\mathbf{x};\mathbf{a};\mathbf{b} \right)\right\rangle  \frac{1}{\Gamma}\left( \left\langle -\frac{\partial U}{\partial \mathbf{x}}\left(\mathbf{x};\mathbf{a};\mathbf{b} \right)\right\rangle+\mathbf{f}\right)\cdot dt\, \, .
\en

We can write the equations  obtained in terms of the variable $\mathbf{X}$:

\be
\mathbf{X}\equiv\mathbf{x}- \mathbf{a}(t) = \mathbf{x}-\mathbf{v} . t \, \, \, \, ,
\en
thinking the potential $U$ as $U(\mathbf{x}(t)-\mathbf{a}(t); \mathbf{b}(t))=U(\mathbf{X}(t); \mathbf{b}(t))$.

It is easy to verify that:

\be \label{a}
\frac{\partial U}{\partial \mathbf{X}}\left(\mathbf{X};\mathbf{b} \right)=\frac{\partial U}{\partial \mathbf{x}}\left(\mathbf{x}-\mathbf{a};\mathbf{b} \right) \, \, \, \, ,
\en

\be \label{c}
\frac{\partial U}{\partial \mathbf{b}}\left(\mathbf{X};\mathbf{b} \right)=\frac{\partial U}{\partial \mathbf{b}}\left(\mathbf{x}-\mathbf{a};\mathbf{b} \right) \, \, \, \, ,
\en

and then we have from Eq. (\ref{trabalho2}), reordering terms

\be \label{trabalho3}
\left\langle d'W\right\rangle-\frac{1}{\Gamma}\left\langle \frac{\partial U}{\partial \mathbf{X}}\left(\mathbf{X};\mathbf{b} \right)\right\rangle^{2}dt + \frac{1}{\Gamma}  \left( \left\langle \frac{\partial U}{\partial \mathbf{X}}\left(\mathbf {X};\mathbf{b} \right)\right\rangle \mathbf{f} \right)\cdot dt=\left\langle \frac{\partial U}{\partial \mathbf{b}}\left(\mathbf{X};\mathbf{b} \right)\right\rangle\cdot d\mathbf{b}\, \, .
\en 

or, in a more compact form by defining $\mathbf{\phi}\equiv\mathbf{f}-\Gamma \mathbf{v}$

\be \label{trabalho3bis}
\left\langle d'W\right\rangle + \mathbf{\phi} \mathbf{v}dt= \left\langle \frac{\partial U}{\partial \mathbf{b}}\left(\mathbf{X};\mathbf{b} \right)\right\rangle\cdot d\mathbf{b}\, \, .
\en

We are going to  write the last equation in terms of the  non-equilibrium free energy, $F^{*}$, defined in Eq. (\ref{helno}).  Using the "` Ehrenfest type"' identity:

\be \label{helno12}
 \frac{\partial F^{*}}{\partial \mathbf{b}}d\mathbf{b}=\left\langle\frac{\partial U}{\partial \mathbf{b}}\right\rangle d\mathbf{b} \, \, \, ,
\en
and, knowing that at $T=constant$ and $\phi=constant$, the diferential of $F^{*}(T, \phi, \mathbf{b})$ is given by:

\be
dF^{*}=\frac{\partial F^{*}}{\partial b}d\mathbf{b} \, \, \, ,
\en
or, using (\ref{helno12})

\be \label{dfree}
dF^{*}=\left\langle\frac{\partial U}{\partial \mathbf{b}}\right\rangle d\mathbf{b} \, \, \, ,
\en

then we have for (\ref{trabalho3bis})

\be \label{trabalho4}
\left\langle d'W\right\rangle + \mathbf{\phi} \mathbf{v}dt= dF^{*}\, \, .
\en

where we define $d'W_{HKW} \equiv -\mathbf{\phi} \mathbf{v}dt$ as the house-keeping work (HKW) needed for a macroscopic ensemble of the stochastic system to keep its instantaneous state as a steady state.

Eq. \ref{trabalho4} can be written explicitly

\be \label{trabalho5}
\left\langle d'W\right\rangle-\frac{1}{\Gamma}\left\langle \frac{\partial U}{\partial \mathbf{X}}\left(\mathbf{X};\mathbf{b} \right)\right\rangle^{2}dt + \frac{1}{\Gamma}  \left( \left\langle \frac{\partial U}{\partial \mathbf{X}}\left(\mathbf {X};\mathbf{b} \right)\right\rangle \mathbf{f} \right)\cdot dt=dF^{*}\, \, .
\en 

Note that, for the case $\mathbf{f}=0$, this equation reduces to (5.3) of Reference \cite{ken2}, proposed there for QS isothermal processes, but with $F^{*}$ instead of $G^{*}$ (Legendre transformation of $F^{*}$). \footnote{In appendix B we show that if this result is expressed in terms of a new free energy G*, as was proposed in Ref\cite{ken2}, new terms should appear.}

The free energy variation  $dF^{*}$  only includes the energy variations due to the change of the parameters $\mathbf{b}$ in potential $U(\mathbf{X},\mathbf{b})$ (see (\ref{dfree})), but not due to the external field $ \phi $ (as should be with G*, see Apendix)\cite{private}. Thus, $dF^{*}$ is related only to the work due to the change of $\mathbf{b}$ (Eq.(\ref{dfree})). As such, the free energy $F^{*}$  is best to analyze a model where the control parameter is $\mathbf{b}$ and the field $\phi$ remains constant.

\subsection{The work $\left\langle  W \right\rangle$ in coordinates  $\mathbf{X}$.}
\label{21}

Using (\ref{a}) and (\ref{c}) the elemental work, Eq. (\ref{trabalho}), can be written as

\be \label{trabalhoX}
d'W=\frac{\partial U}{\partial \mathbf{b}}\left(\mathbf{X}(t);\mathbf{b}(t) \right)\cdot d\mathbf{b}-\frac{\partial U}{\partial \mathbf{X}}\left(\mathbf{X}(t);\mathbf{b}(t) \right)\cdot d\mathbf{a}
\en

If the control parameters $\mathbf{b}$ change from $\mathbf{b}(0)\equiv\mathbf{b}_i$ to $\mathbf{b}(\Delta t)\equiv\mathbf{b}_f$, then the total work $\left\langle W\right\rangle$ performed on the system along a particular process $\mathbf{X}(t) \left(  0 \leq  t \leq \Delta t \right)$ is given by

\be \label{w0}
 W= \int^{\Delta t}_{0}dt \frac{\partial U}{\partial \mathbf{b}}(\mathbf{X}(t); \mathbf{b}(t))\cdot\frac{d\mathbf{b}(t)}{dt}-\int \frac{\partial U}{\partial \mathbf{X}}\left(\mathbf{X}(t);\mathbf{b}(t) \right)  \mathbf{v}dt\, \, \, \, ,
\en

The ensemble average of the work, $ \langle W  \rangle$, over a possible realization of $\left\{\mathbf{\xi}(t)\right\}_{0\leq t \leq \Delta t}$ can be computed as

\be \label{wst}
\langle W  \rangle + \int^{\Delta t}_{0}dt \int \mathbf{dX}P(\mathbf{X},t)\frac{\partial U}{\partial \mathbf{X}}\left(\mathbf{X}(t);\mathbf{b}(t) \right)\mathbf{v}= \int^{\Delta t}_{0}dt \left[\int d\mathbf{X} P(\mathbf{X},t)\frac{\partial U}{\partial \mathbf{b}}(\mathbf{X}; \mathbf{b}(t))\right].\frac{d\mathbf{b}(t)}{dt} ,
\en

where $P$ is the probability distribution function of $\mathbf{X}$ that satisfies the  Fokker-Planck equation, which is given by
 
\be \label{fpst}
\frac{\partial P}{\partial t}(\mathbf{X},t)=-{\cal L}_{FP}(\mathbf{b}(t))P(\mathbf{X},t)
\en
where 

\be
{\cal L}_{FP}(\mathbf{b}(t))\equiv\frac{\partial}{\partial \mathbf{X}}\cdot\Gamma^{-1}\cdot {}^{{}^t}\left( \frac{\partial U}{\partial \mathbf{X}}(\mathbf{X}; \mathbf{b}(t))+\Gamma \mathbf{v}-\mathbf{f}+k_{B}T \frac{\partial }{\partial \mathbf{X}} \right) \, .
\en

Using that 

\be
\langle \frac{\partial U}{\partial \mathbf{X}}(\mathbf{X}; \mathbf{b}(t)) \rangle =\mathbf{f}- \Gamma \mathbf{v} \equiv \phi 
\en

we have for (\ref{wst})

\be \label{wst2}
\langle W  \rangle + \int^{\Delta t}_{0}dt \phi \mathbf{v}= \int^{\Delta t}_{0}dt \left[\int d\mathbf{X} P(\mathbf{X},t)\frac{\partial U}{\partial \mathbf{b}}(\mathbf{X}; \mathbf{b}(t))\right].\frac{d\mathbf{b}(t)}{dt} \,\,\,.
\en

Having computed the spatial integral\footnote{See the appendix where, in order  not to deviate
the text from the principal line of reasoning, all the computations are provided.} in Eq.(\ref{wst2}) we obtain, for long times, $\Delta t$:

\be \label{trabajo}
\langle W \rangle + \int^{\Delta t}_{0}dt \phi \mathbf{v}=\Delta F^{*} + \frac{1}{\Delta t}\int^{1}_{0}ds \frac{d\hat{\mathbf{b}}(s)}{ds} \Lambda(\mathbf{b})\frac{d\hat{\mathbf{b}}(s)}{ds} +{\cal{O}}(\Delta t^{-2} )\;  , 
\en

where $F^{*}$ is the  non-equilibrium free energy and $\Lambda(\mathbf{b})$ is a positive definite $n \times n$ matrix, both defined in the appendix, Eqs. (\ref{helno}) and (\ref{lambdast}) respectively.

Note that the second term in the LHS of Eq. (\ref{trabajo}) represents the house keeping work: $W_{HKW}=-\int^{\Delta t}_{0}dt \phi \mathbf{v}$ . Then, in the limit of slow and smooth change of external parameters ($\Delta t \rightarrow\infty$), Eq. (\ref{trabajo}) tells us that the stochastic energetics gives the correct thermodynamical result of quasi-steady-isothermal processes obtained by Hatano-Sasa \cite{hatano, trepagnier}: $\langle W \rangle-W_{HKW}=\Delta F^{*}$. A discussion on this issue is presented in the section \ref{5}.

\subsection{ A complementarity relation.}
\label{22}
If we define the net work as 

\be\label{network}
\langle {\cal W} \rangle\equiv \langle W \rangle + \int^{\Delta t}_{0}dt \phi \mathbf{v}=\langle W \rangle-W_{HKW},
\en

then the total irreversible work, $\langle {\cal W} \rangle-\Delta F^{*}$, for a very slow process  is given, from Eq. (\ref{trabajo}), by

\be\label{uncert}
\langle {\cal W} \rangle-\Delta F^{*}\approx \frac{1}{\Delta t}\int^{1}_{0}ds \frac{d\hat{\mathbf{b}}(s)}{ds} \Lambda(\mathbf{b})\frac{d\hat{\mathbf{b}}(s)}{ds} \;  , \; \;\; \;\Delta t \rightarrow\infty \;\; .
\en

The integral on the r.h.s of Eq.(\ref{uncert}) has the form of a classical action for a particle of ``mass`` $\Lambda(\mathbf{b})$ and has a minimum ${\cal S}_{min}(\mathbf{c}_{i},\mathbf{b}_{f})$ for a certain ``classical `` path. Hence, as in the case of  quasi-equilibrium process, an inequality that resembles a sort of  `` uncertainty'' relation remains true  for the present case of steady process, valid asymptotically for $\Delta t \rightarrow\infty$:

\be\label{uncert2}
\left(\langle {\cal W}  \rangle-\Delta F^{*}\right)\Delta t\geq{\cal S}_{min}(\mathbf{b}_{i},\mathbf{b}_{f})\; \;.
\en

According to (\ref{uncert2}), the estimation of the non-equilibrium Helmholtz free energy, by the measurement of the net mean mechanical work, contains an indetermination $ Q_{irr}= \langle {\cal W} \rangle - \Delta F^{*}$ (the total irreversible work), whose product by $\Delta t$  cannot be smaller than  a positive lower bound. The precise determination of the non-equilibrium Helmholtz free energy through the observation of the work $\langle W' \rangle$ requires an indefinitely large experimental time $\Delta t$.

\subsection{The second principle.}
\label{23}

We can express our results for an elementary process. From (\ref{uncert}), we have up to the first order

\be \label{dexp6}
\langle d'{\cal W}  \rangle   =  dF^{*}  +  \frac{d\hat{\mathbf{b}}}{dt} \Lambda(\mathbf{b})\frac{d\hat{\mathbf{b}}}{dt} dt \, \, ,
\en
where we used (\ref{scaled}) to return  to variable $t$, or explicitly

\be
\left\langle d'W\right\rangle-\frac{1}{\Gamma}\left\langle \frac{\partial U}{\partial \mathbf{X}}\left(\mathbf{X};\mathbf{b} \right)\right\rangle \left\{\left\langle \frac{\partial U}{\partial \mathbf{X}}\left(\mathbf{X};\mathbf{b} \right)\right\rangle-  \mathbf{f}    \right\} dt =dF^{*}  +  \frac{d\hat{\mathbf{b}}}{dt} \Lambda(\mathbf{b})\frac{d\hat{\mathbf{b}}}{dt} dt \,\, .
\en

Equation (\ref{dexp6}) represents the 2nd law for QS processes. Comparing  with equation (5-3), of Ref. \cite{ken2} we note that we were able to obtain it with a 1st order correction\footnote{With the clarifications made after equation (\ref{trabalho7}).}. 

Below we will apply our analysis to a simple  model and, after that, we will discuss  possible applications to  more relevant experiments, i.e. the stretching of a RNA molecule and the drag of a  dipolar particle in a non-uniform electrical field.

\section{Application to a simple model.}
\label{3}

\begin{figure}
\begin{center}
\fbox{
\includegraphics[height=40mm,width=80mm,angle=0]{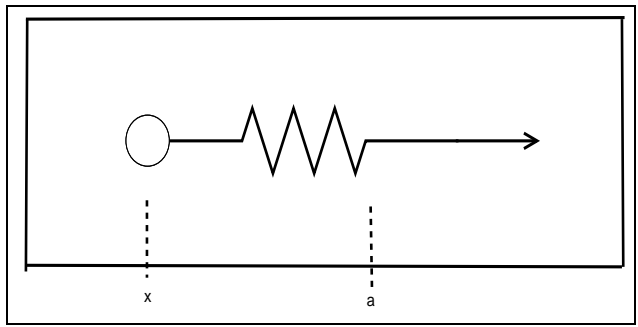}}
\caption{A particle submitted  to a spring,  immersed in a heath bath.}
\label{figurita}
\end{center}
\end{figure}

As an application of the preceding approach for steady processes, we consider a single particle trapped in an one-dimensional harmonic potential  $U(x-a(t), b(t))=\frac{1}{2} b(t) (x-a)^{2}$ (i.e. a spring) and immersed in a heath bath at temperature $T$, being the strength $b(t)$ the control parameter that changes slowly during a time lapse (see Fig. \ref{figurita}).
The variable $x$ represents the displacement of the particle from the origin and $a(t)$ may be regarded, up to a constant difference, as the position of the opposite end of the spring. We study this system under the change of the {\it drive} $v(t)\equiv\frac{da(t)}{dt}$\footnote{This is the example presented in Ref. \cite{ken2}, section 5.}
In terms of the variable $X\equiv x-a$ we can write the potential as $U(X(t),b(t))=\frac{1}{2} b(t) X^{2}$.

We were able to compute the quantity $\Lambda(\mathbf{b})$  from its simplified formula Eq (\ref{lambdast2}), now that (one-dimension)  is a $1 \times 1$ matrix, i.e. a scalar quantity. In this model, the steady distribution (recall $v=constant$), in the absence of an external force ($\mathbf{f}=0$), is given by

\be \label{lostmodel}
P_{st}(X;b,v)= \frac{e^{-\beta(\frac{1}{2} b(t) X^{2}+ \Gamma v \cdot X) }}{\int dX e^{-\beta(\frac{1}{2} b(t)X^{2}+\Gamma v \cdot X)}} \, \, ,
\en

and the kernel $g(X,X';b)$ is

\be
g(X,X';b)=\frac{\beta\Gamma}{2}sgn(X-X')\int_{X'}^{X}\frac{1}{P_{st}(X;b,v)}\, \, .
\en

After integration of (\ref{lambdast2}), we found:

\begin{eqnarray}\label{numeritos}
\Lambda(b) = \frac{\Gamma}{4 \beta b^3} + \frac{\Gamma^3 v^2}{b^4} \, \, ,
\end{eqnarray}

thus, from (\ref{dexp6}), we have for the irreversible heat:

\be \label{calorcito}
\langle d{\cal W}  \rangle -  dF^{*} \equiv dQ_{irr} =    \left(\frac{\Gamma}{4 \beta b^3} + \frac{\Gamma^3 v^2}{b^4} \right) \dot{b}^{2} dt \, \, 
\en

We see that two quite different terms contribute to $\Lambda(b)$, and therefore to the irreversible heat released during the whole process of variation of the parameter $b$.
By means of qualitative analysis, we are going to show that the difference between these two terms lies in their physical origin. 
The first term depends directly on temperature, and contains the information about the effect of the change of $b$ on the mean fluctuations of the particle position, when it is coupled to the thermal bath.
The second term is the only one with a contribution due to the steady regime, that is, depends on the velocity $v$, and is due to the rearrangement of the equilibrium position $X_{eq}$.\\   
In thermal equilibrium, 
 
\be
\frac{1}{2}k_{B}T = <U(X,b)> = \frac{1}{2}b <X^{2}>\,\, ,
\en

thus, we can define a typical mean displacement (or equivalently, the typical mean amplitude of oscillation) as:

\be
X_{d} \equiv \sqrt{\frac{k_{B}T}{b}} \, \, 
\en

with an equilibrium position that in the steady regime is

\be
X_{eq}=-\frac{\Gamma v}{b} \, \, .
\en

We associate $\dot{X_{d}}$ and   $\dot{X}_{eq}$ with typical velocities of adjustment, $v_1$ $v_2$, due to the change of the parameter $b$. They are given by
\be
v_1\equiv\dot{X_{d}}=-\frac{1}{2}\sqrt{k_{B}T}\frac{\dot{b}}{b^{\frac{3}{2}}} \, \, 
\en
and

\be
v_2\equiv\dot{X}_{eq}=\frac{\Gamma v}{b^2}\dot{b} \, \, .
\en

Then we can estimate the irreversible heat as the energy dissipated  by the friction force, $\Gamma v_{i}$, ($i=1,2$), in the viscous medium, in the infinitesimal lapse $dt$. It is given by 
\be\label{calorir}
dQ_{irr} =\Gamma v_{1}^2dt + \Gamma v_{2}^2dt=\left(\frac{\Gamma}{4 \beta b^3} + \frac{\Gamma^3 v^2}{b^4} \right) \dot{b}^{2} dt \, \, 
\en
which is exactly Eq. (\ref{calorcito}).

It is important to note that for the case $v=0$ (quasi-equilibrium) we re-obtain the result found for $\Lambda$ in Eq.(18) of Ref.\cite{ken1}. \footnote{The present potential  was studied in \cite{ken1}, Remark 4, where the stiffness constant is named $a$ instead of $b$.}
When $\Lambda$ is replaced in expression Eq. (\ref{dexp6}) and is integrated during the time lapse, the integral can be minimalized in order to obtain an optimal protocol for which the irreversible heat is minimum \cite{ken1}. However, care is required in calculating the optimum protocol as control parameters usually must undergo jumps in the extremes (initial and/or final) of the time interval of operation. This fact is well know in the community of control theory, but it was first introduced in the domain of SE by Siefert \cite {Seif}.

\section{Stretching a RNA molecule in a steady configuration.}
\label{4}

\begin{figure}[t]
\begin{center}
\fbox{
\includegraphics[height=60mm,width=80mm,angle=0]{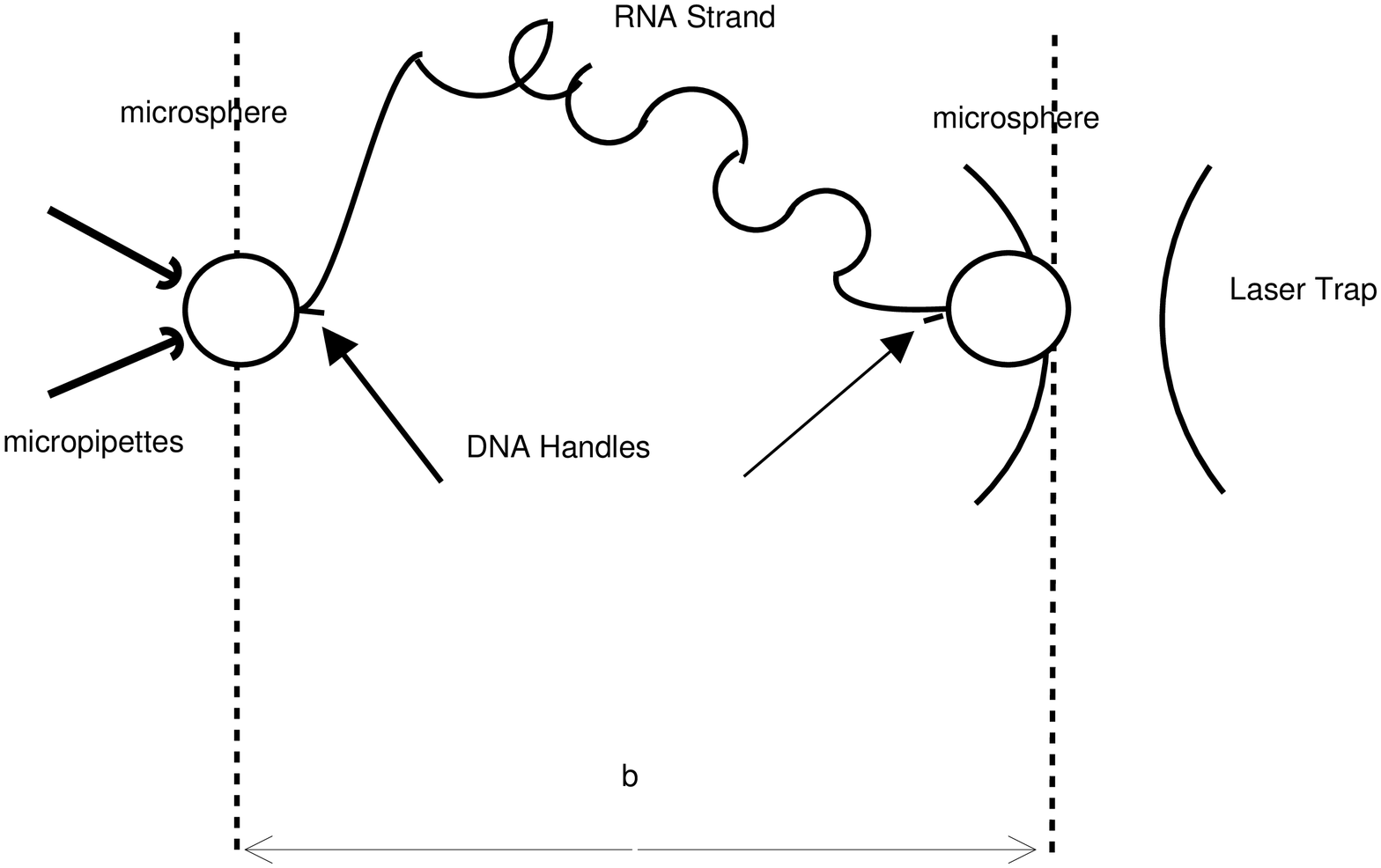}}
\caption{Stretching a RNA molecule}
\label{RNA}
\end{center}
\end{figure}

As a possible more realistic and relevant aplication of the results obtained, we consider an experiment consisting of a strand of RNA with two microspheres glued to its extremes by mean of DNA handles. One microsphere is held in place by a micropipette and the other is confined by an optical trap. The system is immersed in a solution at certain temperature and it is possible to control de distance from the end of the micropipette  to the optical trap. The optical trap creates an approximately harmonic potential for the microsphere, playing the role of the spring in the simple model of the last section (Fig. \ref{RNA}).
\footnote{This is  analog to  the experiment presented in section 3 of \cite{Jarzinsky2008} but here we consider a steady situation, i.e. for example, the bath is a steady flow.}. In this experiment work ($W$)  is performed on the RNA strand as we strech it by varying the parameter $b$ from $b_{i}$ to $b_{f}$ following a protocole between the initial ($i$) and final ($f$) steady states. By repeating the same experiment many times between the same steady states we can construct the average work $\langle W \rangle $ and, after substracting the house-keeping work, we could obtain the "`net work"' 
$ \langle {\cal W} \rangle$ that satisfies, for a large but finite $\Delta t$, the equation (\ref{uncert2}).

\section{Dipolar particle in the presence of an  electric or  magnetic field}
\label{4}

Another possible aplication we propose is given by a dipolar particle\footnote{It may be a particle on which  a dipole  is induced. We consider for simplicity an unidimensional model.} dragged through an aqueous solution and subjected to a non-uniform electrical (or magnetic \cite{wirtz}) field, which gradient $\frac{\partial\mathbf{ E}}{\partial X }$ produces a force acting on the dipolar particle. The particle is driven by this force across a potential\cite{speck} which, to a first approximation, is supposed to be harmonic. Its stiffness is time dependent, in such a way that the particle moves at constant velocity. By varying the gradient, as an external parameter $c\equiv \frac{\partial\mathbf{ E}}{\partial X }$ following a protocole, work is performed on the dipolar particle. Suppose the system is prepared in a steady state at $c=c_{i}$ then is performed work on the dipolar particle by varying $c$ from $c_{i}$ to $c_{f}$ according to certain protocole and finally the system is allowed to reach again a steady state with the value of the parameter given by $c_{f}$.

We were able to compute the quantity $\Lambda(\mathbf{c})$  from its simplified formula Eq (\ref{lambdast2}). In this model, the steady distribution (recall $v=constant$) is given by

\be \label{dipolemodel}
P_{st}(X;c,v)= \frac{e^{-\beta(\frac{1}{2} b(t) X^{2}- D \cdot c \cdot X + \Gamma v \cdot X) }}{\int dX e^{-\beta(\frac{1}{2} b(t) X^{2}-D \cdot c \cdot X+\Gamma v \cdot X)}} \, \, ,
\en

and the kernel $g(X,X';c)$ is

\be
g(X,X';c)=\frac{\beta\Gamma}{2}sgn(X-X')\int_{X'}^{X}\frac{1}{P_{st}(X;c,v)}\, \, .
\en

After integration of (\ref{lambdast2}), we found:

\begin{eqnarray}\label{numeritosdipole}
\Lambda(c) = \frac{\Gamma D^2}{b^2}  \, \, ,
\end{eqnarray}

thus, from (\ref{dexp6}), we have for the elementary irreversible heat:

\be \label{calorcitodipole}
\langle d{\cal W}  \rangle -  dF^{*} \equiv dQ_{irr} =    \left(\frac{\Gamma D^2}{ b^2} \right) \dot{c}^{2} dt \, \, 
\en

A qualitative analysis allows us to verify that the irreversible heat $dQ_{irr}$, Eq. (\ref{calorcitodipole}), is due to the rearrangement of the equilibrium position $X_{eq}$:

\be
X_{eq}=-\frac{\Gamma v}{b}+\frac{D c}{b} \Rightarrow
\en

\be
\dot{X_{eq}}=\frac{D}{b}\dot{c}
\en

Assimilating $\dot{X_{eq}}$ to  a characteristic velocity of the process, we have:

\be \label{calorcitodipole2}
 dQ_{irr} = \Gamma (\dot{X_{eq}})^2 dt =  \left(\frac{\Gamma D^2}{ b^2} \right) \dot{c}^{2} dt \, \, ,
\en

which is in agreement with the result of Eq. (\ref{calorcitodipole}).

The amplitude of the oscillations is not affected by the variation of the parameter $c$, therefore a term dependent on the temperature in equation (\ref{calorcitodipole}) does not appear  as it happens when  the parameter $b$ is varied, Eq.(\ref{calorir}). Also note that, in this example, $ dQ_{irr}$ does not depends on the velocity. These differences (shown by Eq. (\ref{calorir}) and Eq. (\ref{calorcitodipole})  ) would allow an experimental verification of the model.

\section{Conclusion}
\label{5}
Following an analogue approach  for the case of the quasi-equilibrium processes we were able so show that an inequality, connecting the irreversible net work $\langle {\cal W}   \rangle $  and the experimental time $\Delta t$, exists for the case of QS processes. It is given by (\ref{uncert2}) and it states that the estimation of the non-equilibrium Helmholtz free energy, by the measurement of the net mechanical work, contains an indetermination  $ Q_{irr}= \langle {\cal W}  \rangle - \Delta F^{*}$ (the total irreversible work), whose product by $\Delta t$  cannot be smaller than a positive lower bound. The precise determination of the non-equilibrium Helmholtz free energy through the observation of the work $\langle {\cal W} \rangle$ requires an indefinitely large experimental time $\Delta t$.
We deduced the second law for QS processes with a 1st order correction for a particle subject to a potential and immersed in a heath bath and we showed that the $HKW$ appears to be naturally subtracting from the mean work.

We have discussed the applicability of the results to two experiments: i)  a RNA molecule is stretched by an optical trap and micropipettes; ii) a dipolar particle immersed in an aqueous solution and subjected  to a non-uniform electrical field. Both experimental arrangements were designed so that initial and end states are stationaries. The results obtained indicate a way to test the model experimentally, as explained in section 5.

Finally \footnote{We appreciate the suggestion of this idea made by the anonymous referee 2.} it is very interesting to discuss the role that the result obtained in our work Eq. (\ref{uncert2}) could play in the context of the Hatano-Sasa equation  (\cite{hatano} Eq.(8)), which is a kind of generalization of the Jarzinsky equation. It is well known that Jarzinsky equation 

\be
  \left\langle e^{-\beta W}\right\rangle =e^{-\beta\Delta F}
\en

gives us (using Jensen's inequality\footnote{$\langle e^{x} \rangle  \geq e^{\langle x \rangle} \, .$}) an expression of the 2nd principle that applies to the processes of non-quasi-equilibrium

\be \label{2nd}
\left\langle  W \right\rangle - \Delta F \geq 0. 
\en
This expression can be viewed as a complementary aspect of the result ("`uncertainty relation"')  that gives (to first order in $1/\Delta t $) a minimum (positive) value for the product  $\left\langle  W \right\rangle - \Delta F $, obtained by Sekimoto and Sasa (\cite{ken1} Eq. (11)) 

\be
\left\langle  W \right\rangle - \Delta F \geq \frac{k_{B}T}{\Delta t} {\cal S}_{min}
\en

in the sense that Eq. (\ref{2nd}) is valid for all orders although it does not provide the value (other than zero).

Similarly from equation (8) of Hatano-Sasa \cite{hatano} (which represents a  generalization of  Jarzinsky equation,  valid for non-quasi-stationary processes)  it is possible to derive a minimum work principle for these processes, Eq. (25) of \cite{hatano}:

\be
\left\langle  W_{ex} \right\rangle - \Delta F \geq 0 
\en

and in the same way we might think that this equation can be viewed as a complementary aspect of 
the result we have obtained in Eq. (\ref{uncert2})
\footnote{Note that the work $ \langle {\cal W}  \rangle $ in Eq. (\ref{uncert2}) and $ \langle W_{ex}\rangle $ play a similar role as both have discounted the housekeeping work.}.

\section{Acknowledgments}
The authors wish to thank Prof. Ken Sekimoto  for the fruitfull discussions and private communications. We also wish to thank the enlightened criticism made by the anonymous referee 2.
We would  like to thank Consejo Nacional de Investigaciones Cient\'{\i}ficas y T\'ecnicas from Argentina and the Minist\'erio da Ci\^encia, Tecnologia e Inova\c c\~ao/ CNEN / CBPF of Brazil for their financial support. 

\vspace{1cm}

\appendix

\section{Proof of Eq. (\ref{trabajo}):  the work in coordinates $\mathbf{X}$.}

In order to compute the integral from (\ref{wst}), the scaled time $s$

 \be\label{scaled}
 s\equiv\frac{t}{\Delta t}
 \en 
 
 is defined. The probability distribution depending on this argument is defined as
 
 $\hat{P}(\mathbf{X},s;\Delta t)\equiv P(\mathbf{X}, s\Delta t)$,
  
 and the parameters as $\hat{\mathbf{b}}(s)\equiv\mathbf{b}(s \Delta t) $. Equations (\ref{wst}) and (\ref{fpst}) become

\be \label{wsts}
\langle W  \rangle + \int^{\Delta t}_{0}dt \langle  \frac{\partial U}{\partial \mathbf{X}}\left(\mathbf{X}(t);\mathbf{b}(t) \right)\rangle \mathbf{v}= \int^{1}_{0}ds \frac{d\hat{\mathbf{b}}(s)}{ds} \cdot \int d\mathbf{X} \frac{\partial U}{\partial \mathbf{b}}(\mathbf{X}; \hat{\mathbf{b}}(s))\hat{P}(\mathbf{X},s;\Delta t)  ,
\en

\be \label{fpsts}
\frac{1}{\Delta t}\frac{\partial \hat{P}}{\partial s}(\mathbf{X},s;\Delta t)=-{\cal L}_{FP}(\hat{\mathbf{b}}(s))\hat{P}(\mathbf{X},s;\Delta t)\ .
\en

Eq. (\ref{fpsts}) can be solved perturbatively  by assuming that $\Delta t$ is large enough to make an expansion of $P$ in powers of $\frac{1}{\Delta t}$ as

\be\label{perst}
\hat{P}(\mathbf{X},s;\Delta t)= \hat{P}^{(0)}(\mathbf{X},s)+\frac{1}{\Delta t}\hat{P}^{(1)}(\mathbf{X},s)+\cdots\ .
\en
Substituting in (\ref{fpsts}), we have for the zero and  first order

\be \label{0st} 
0=-{\cal L}_{FP}(\hat{\mathbf{b}}(s))\hat{P}^{(0)}(\mathbf{X},s),\,\,\,\,\,\,\,0^{th} order 
\en

\be \label{1st}
\frac{\partial \hat{P}^{(0)}}{\partial s}(\mathbf{X},s)=-{\cal L}_{FP}(\hat{\mathbf{b}}(s))\hat{P}^{(1)}(\mathbf{X},s)\,\,\,\,\,\,\,1^{st} order .
\en

From the lowest order, Eq. (\ref{0st}), and the normalization condition $ \int d\mathbf{X}\hat{P}^{(0)}(\mathbf{X},s)=1$ we deduce that 
 $\hat{P}^{(0)}$ is the {\it steady distribution} $P_{st}$ for a given parameter $\hat{\mathbf{b}}(s)$:

\be \label{lost}
\hat{P}^{(0)}(\mathbf{X},s) = P_{st}(\mathbf{X};\hat{\mathbf{b}}(s),\mathbf{v})\equiv \frac{e^{-\beta(U(\mathbf{X},\mathbf{b})- \phi\cdot\mathbf{X}) }}{\int d\mathbf{X}e^{-\beta(U(\mathbf{X},\mathbf{b})+ \phi\cdot\mathbf{X}) }} \, \, 
\en
where $\beta\equiv\frac{1}{k_{B}T}$ and $\phi \equiv \mathbf{f}-\Gamma \mathbf{v} = constant$ ( i.e. : $\mathbf{f}$ and $\mathbf{v}$ are  constants ).

Eq. (\ref{1st}) becomes

\be \label{1steady}
\frac{\partial P_{st}}{\partial s}(\mathbf{X};\hat{\mathbf{b}}(s),\mathbf{v})=-{\cal L}_{FP}(\hat{\mathbf{b}}(s))\hat{P}^{(1)}(\mathbf{X},s)\, \, .
\en

Now, the kernel $g(\mathbf{X},\mathbf{X}';\hat{\mathbf{b}}(s))$ is defined as the solution of

\be\label{kerst}
-{\cal L}_{FP}(\mathbf{b})\left[P_{st}(\mathbf{X};\hat{\mathbf{b}}(s)) g(\mathbf{X},\mathbf{X}';\hat{\mathbf{b}}(s))\right]=\delta(\mathbf{X},\mathbf{X}')\; .
\en

If we multiply  Eq.(\ref{1steady}) by $P_{st}(\mathbf{X};\hat{\mathbf{b}}(s)) g(\mathbf{X},\mathbf{X}';\hat{\mathbf{b}}(s),\mathbf{v})$ and then integrate in $\mathbf{X} $, we obtain $\hat{P}^{(1)}(\mathbf{X},s)$ as

\be \label{p1}
\hat{P}^{(1)}(\mathbf{X},s)= P_{st}(\partial s)(\mathbf{X};\hat{\mathbf{b}}(s),\mathbf{v})\left[ \int d\mathbf{X}' g(\mathbf{X},\mathbf{X}';\hat{\mathbf{b}}(s)) \frac{\partial P_{st}}{\partial s}(\mathbf{X}';\hat{\mathbf{b}}(s),\mathbf{v}) + \chi  \right] \; ,
\en
where the integration constant $\chi $ is obtained  from the normalization condition, 

$\int d\mathbf{X} \hat{P}^{(1)}(\mathbf{X},s)=0$, as

\be \label{chi}
\chi = -\int d\mathbf{x}\left\{P_{st}(\mathbf{X};\hat{\mathbf{b}}(s))\int d\mathbf{X}' g(\mathbf{X},\mathbf{X}';\hat{\mathbf{b}}(s)) \frac{\partial P_{st}}{\partial s}(\mathbf{X}';\hat{\mathbf{b}}(s))\right\} 
\en
Having obtained $\hat{P}(\mathbf{X},s)$ up to the first order, we substitute  

$\hat{P}(\mathbf{X},s)=P_{st}(\mathbf{X};\hat{\mathbf{b}}(s),\mathbf{X})+\frac{1}{\Delta t}\hat{P}^{(1)}(\mathbf{X},s)+...$ in Eq. (\ref{wsts}) and we have

\be \label{wsts1}
\langle W  \rangle + \int^{\Delta t}_{0}dt \langle  \frac{\partial U}{\partial \mathbf{X}}\left(\mathbf{X}(t);\mathbf{b}(t) \right)\rangle \mathbf{v}= \int^{1}_{0}ds \frac{d\hat{\mathbf{b}}(s)}{ds} \cdot \int d\mathbf{X} \langle\frac{\partial U }{\partial \mathbf{b}}(\mathbf{X}; \hat{\mathbf{b}}(s))\rangle \left\{P_{st}(\mathbf{X};\hat{\mathbf{b}}(s))+\frac{1}{\Delta t}\hat{P}^{(1)}(\mathbf{X},s)+...\right\}.
\en

As we are dealing with an out of equilibrium process, it is useful to make use of the
 non-equilibrium Helmholtz free energy $F^{*}(T,\mathbf{b}, \mathbf{v})$, defined by Sekimoto in \cite{ken2}, that is given by\footnote{We can obtain  $F^{*}(T,\phi,\mathbf{b})$ in an operational way by mean of the equation (\ref{trabajo}).}

\be \label{helno}
F^{*}(T,\phi,\mathbf{b})\equiv -k_{B} T \ln\left[{\int \exp{-\frac{U(\mathbf{X}; \hat{\mathbf{b}})-\phi\cdot \mathbf{X} }{k_{B} T}}d\mathbf{X}}\right] \, \, \, ,
\en

The following "Ehrenfest type"  identity, concerning the steady ensemble average $\left\langle\frac{\partial U}{\partial \mathbf{b}}\right\rangle_{P_{st}}$, is satisfied:

\be \label{helno1}
 \frac{\partial F^{*}}{\partial \mathbf{b}}=\left\langle\frac{\partial U}{\partial \mathbf{b}}\right\rangle_{P_{st}}\equiv \int d\mathbf{X} \frac{\partial U}{\partial \mathbf{b}}(\mathbf{X}; \hat{\mathbf{b}})P_{st}(\mathbf{X};\hat{\mathbf{b}}) \, \, \, ,
\en
and furthermore
\be \label{helno2}
\frac{\partial F^{*}}{\partial \phi}= -\langle \mathbf{X}  \rangle\, \, \, ,
\en

so we have, from Eq.(\ref{wsts1}), up to the first order

\be \label{lavoro1}
\langle W  \rangle + \int^{\Delta t}_{0}dt \langle  \frac{\partial U}{\partial \mathbf{X}}\left(\mathbf{X}(t);\mathbf{b}(t) \right)\rangle \mathbf{v}= \int^{f}_{i}d\mathbf{b}\frac{\partial F^{*}}{\partial \mathbf{b}}+ \frac{1}{\Delta t} \int^{1}_{0}ds \frac{d\hat{\mathbf{b}}(s)}{ds} \cdot \int d\mathbf{X} \frac{\partial h }{\partial \mathbf{b}}(\mathbf{X}; \hat{\mathbf{b}}(s))\hat{P}^{(1)}(\mathbf{X},s)+{\cal{O}}(\Delta t^{-2} )\,\, .
\en

\vspace{0.5cm}

Using the relation ("`chain's rule"')
$\frac{\partial P_{st}}{\partial s}(\mathbf{X}';\hat{\mathbf{b}},\mathbf{v})= {}^t\left(\frac{\partial P_{st}}{\partial \hat{\mathbf{b}}}(\mathbf{X}';\hat{\mathbf{b}}(s),\mathbf{v})\right)\cdot \frac{d\hat{\mathbf{b}}(s)}{ds}$, and substituting (\ref{p1}) (using (\ref{chi})) in the first order term of (\ref{lavoro1}), we  have for $\langle W  \rangle$

\be \label{trabmedio}
\langle W  \rangle + \int^{\Delta t}_{0}dt \langle  \frac{\partial U}{\partial \mathbf{X}}\left(\mathbf{X}(t);\mathbf{b}(t) \right)\rangle \mathbf{v}= \int^{f}_{i}\left(d\mathbf{b}\frac{\partial F^{*}}{\partial \mathbf{b}}+d\phi\frac{\partial F^{*}}{\partial \phi} \right)+
\frac{1}{\Delta t}\int^{1}_{0}ds \frac{d\hat{\mathbf{b}}(s)}{ds} \Lambda(\mathbf{b})\frac{d\hat{\mathbf{b}}(s)}{ds}+ {\cal{O}}(\Delta t^{-2})
\en
where

\begin{eqnarray}\label{lambdast}
 \Lambda(\mathbf{b}) &\equiv& 
   \int d\mathbf{X} \int d\mathbf{X}'\,\,{}^t \left(\frac{\partial h}{\partial \mathbf{b}}(\mathbf{X}; \hat{\mathbf{b}})\right)P_{st}(\mathbf{X};\hat{\mathbf{b}},\mathbf{v}) . \nonumber \\
  & & 
  \int  d\mathbf{\bar{X}}\left(\delta(\mathbf{\bar{X}}-\mathbf{X})-  
P_{st}(\mathbf{\bar{X}};\hat{\mathbf{b}},\mathbf{v})\right)g(\mathbf{\bar{X}},\mathbf{X}';\hat{\mathbf{b}})\left(\frac{\partial P_{st}}{\partial \mathbf{b}}(\mathbf{X}';\hat{\mathbf{b}},\mathbf{v})\right)\,\,  .
\end{eqnarray}
is a positive definite $n \times n$ matrix\footnote{The quantity $\Lambda(\mathbf{b})$ is related with $\Phi$ which is the dissipation function of linear irreversible thermomdynamics for steady states, we have $2\Phi=\int^{1}_{0}ds \frac{d\hat{\mathbf{b}}(s)}{ds} \Lambda(\mathbf{b})\frac{d\hat{\mathbf{b}}(s)}{ds}$. See \cite{ken1} Remark 1 and \cite{landau}.}, which can  be simplified to

\begin{eqnarray}\label{lambdast2}
\Lambda(\mathbf{b})=-\frac{1}{\beta} 
   \int d\mathbf{X} \int d\mathbf{X}'\,\,\frac{\partial P_{st}}{\partial \mathbf{b}}(\mathbf{X};\hat{\mathbf{b}},\mathbf{v}).
 g(\mathbf{X},\mathbf{X}';\hat{\mathbf{b}})\,\, {}^t\left(\frac{\partial P_{st}}{\partial \mathbf{b}}(\mathbf{X}';\hat{\mathbf{b}},\mathbf{v})\right) .
\end{eqnarray}
being $g(\mathbf{X},\mathbf{X}';\hat{\mathbf{b}})$ the  Green's function satisfying Eq.(\ref{kerst}).
\footnote{In order to demonstrate Eq.(\ref{lambdast2}), we note that the operator (distribution) ${\cal R}_{\mathbf{\bar{X}}}^{\bot}(\mathbf{b})$, defined by its action on an arbitrary well behaved function $\psi(\mathbf{X}) $ as ${\cal R}_{\mathbf{X}}^{\bot}(\mathbf{b})\psi(\mathbf{X})\equiv \int  d\mathbf{\bar{X}}\left(\delta(\mathbf{\bar{X}}-\mathbf{X})-  
P_{st}(\mathbf{\bar{X}};\hat{\mathbf{b}},\mathbf{v})\right)\psi(\mathbf{\bar{X}})$, satisfies the following two identities:
$\int d\mathbf{X} P_{st}(\mathbf{\bar{X}};\hat{\mathbf{b}},\mathbf{v})\left[{\cal R}_{\mathbf{X}}^{\bot}(\mathbf{b})\psi(\mathbf{X})\right]=0$ and $\int d\mathbf{X}  \frac{\partial P_{st}}{\partial \mathbf{b}}(\mathbf{X};\hat{\mathbf{b}},\mathbf{v})\left[{\cal R}_{\mathbf{X}}^{\bot}(\mathbf{b})\psi(\mathbf{X})\right]=  \int d\mathbf{X}  \frac{\partial P_{st}}{\partial \mathbf{b}}(\mathbf{X};\hat{\mathbf{b}},\mathbf{v})\psi(\mathbf{X}).$ 
The operator ${\cal R}_{\mathbf{X}}^{\bot}(\mathbf{b})$  is the equivalent for steady states, of the operator defined in \cite{ken1} Eq.(27) for equilibrium states.}.

From the definition of the non-equilibrium Helmholtz free energy, Eq. (\ref{helno}),
for $T=constant$ and $\phi=constant$, we have

\be\label{dF}
dF^{*}=\frac{\partial F^{*}}{\partial \mathbf{b}} \cdot d\mathbf{b}+\frac{\partial F^{*}}{\partial \mathbf{\phi}}\cdot d\mathbf{\phi} = \frac{\partial F^{*}}{\partial \mathbf{b}} \cdot d\mathbf{b} \,\, .
\en

Substituting (\ref{dF}) in (\ref{trabmedio}) and considering $\phi$ constant it follows Eq.(\ref{trabajo}).

\section{G* free energy}

In what follows we are going to show that, using the free energy $G^*$,  an additional term, which was not considered in equation (5.3) of Ref. \cite{ken2},  is introduced.

By means of a Legendre transformation, following Sekimoto\cite{ken2}, we define the new free energy
$ G^{*}(T,\langle\mathbf{X}\rangle,\mathbf{b})$ as

\be \label{Ff3}
G^{*}(T,\langle\mathbf{X}\rangle,\mathbf{b})\equiv F^{*}(T,\phi,\mathbf{b})-\frac{\partial F^{*}(T,\phi, \mathbf{b})}{\partial \phi}\cdot \phi
\en

with
\be
\frac{\partial F^{*}(T,\phi, \mathbf{b})}{\partial \phi}= -\langle\mathbf{X}\rangle \, \, .
\en

We have for the differentials ($\phi = constant$):

\be \label{FG}
dF^{*}=dG^{*}- \phi d\langle\mathbf{X}\rangle \, \, ,
\en

and substituting in Eq (\ref{trabalho4}), we obtain

\be \label{trabalho6}
\left\langle d'W\right\rangle+\phi \mathbf{v}dt+\phi d\langle\mathbf{X}\rangle=dG^{*} \, \, ,
\en

or explicitly

\be \label{trabalho7}
\left\langle d'W\right\rangle-\frac{1}{\Gamma}\left\langle \frac{\partial U}{\partial \mathbf{X}}\left(\mathbf{X};\mathbf{b} \right)\right\rangle \left\{\left\langle \frac{\partial U}{\partial \mathbf{X}}\left(\mathbf{X};\mathbf{b} \right)\right\rangle-  \mathbf{f}    \right\} dt+(\mathbf{f}-\Gamma \mathbf{v}) d\langle\mathbf{X}\rangle=dG^{*} \,\, .
\en 

Comparing this equation with Eq (5.3) of Reference \cite{ken2} we see that an additional term in the LHS is present: 
$ \phi d\langle\mathbf{X}\rangle $ (or $-\Gamma \mathbf{v} d\langle\mathbf{X}\rangle\ $, for the case $\mathbf{f}=0$)\cite{private}.
This term appears because the change of the mean value $\langle\mathbf{X}\rangle\ $ in the field $\phi$ and it is the work necessary for doing that. Accordingly, the free energy variation $dG^{*}$ includes the free energy variations due to the change of the parameters $\mathbf{b}$ in potential $U(\mathbf{X},\mathbf{b})$ 
and the variations due to the change in the mean value of $\langle \mathbf{X} \rangle$ in the field  $\phi$. This can be emphasized writing $dG^{*}$ according to Eq.(\ref{dfree}) and (\ref{FG}) ,

\be \label{G}
dG^{*}=\left\langle\frac{\partial U}{\partial \mathbf{b}}\right\rangle d\mathbf{b}+ \phi d\langle\mathbf{X}\rangle \, \, .
\en

\end{document}